\documentclass[final]{aipproc}
\usepackage{amsmath,amssymb,bbm,mathrsfs}
\usepackage{axodraw,wrapfig}

\layoutstyle{6x9}

\newcommand{\gtr}{\mbox{$\mathsf{SU(3)_C \times SU(3)_L \times SU(3)_R
      \times \mathbbm{Z}_3}$}}  
\newcommand{\Le}{\text{\sc l}}  
\newcommand{\qu}{\text{\sc q}}  
\newcommand{\VEV}[1]{\left\langle #1\right\rangle}        

\begin{document}

\title{Unification without low-energy supersymmetry}



\author{S\"oren Wiesenfeldt}{%
  address={Department of Physics, University of Illinois at
    Urbana-Champaign, 
    \\
    1110 West Green Street, Urbana, IL 61801, USA}%
}

\begin{abstract}
  Without supersymmetry, the gauge couplings in the standard model
  with five Higgs doublets unify around $10^{14}$ GeV.  In this case,
  the trinified model, $\gtr$, with the minimal Higgs sector required
  for symmetry breaking, is the ideal grand-unified candidate.  Small
  neutrino masses are generated via a radiative seesaw mechanism,
  without the need for intermediate scales, additional Higgs fields,
  or higher-dimensional operators.  The proton lifetime is above the
  experimental limits, with the decay modes $p\to\bar\nu K^+$ and
  $p\to\mu^+ K^0$ potentially observable.  The split-SUSY version of
  the model, with one light Higgs doublet, is equally attractive.
\end{abstract}


\maketitle

Grand unification of the strong, weak, and electromagnetic
interactions into a simple gauge group is a very appealing idea that
has been vigorously pursued for many years.  The fact that the three
gauge couplings of the minimal supersymmetric standard model meet
almost exactly at $M_\text{U} = 2\times 10^{16}$ GeV, has made
supersymmetric GUTs a beautiful framework for theories beyond the
standard model (SM).  However, if we go beyond SU(5), as suggested by
neutrino experiments, we do not need single-step unification.
Furthermore, unification is possible without supersymmetry if we
extend the Higgs sector of the SM to five or six Higgs doublets
\cite{unification-higgs}.

In this talk, we present a viable candidate for a unified theory
without (low-energy) supersymmetry \cite{Sayre:2006ma}.  Since the
unification of the gauge couplings occurs at a lower scale,
$M_\text{U}\simeq 10^{14}$ GeV, we do not choose a simple GUT group,
which would yield too rapid proton decay.  Instead, we consider a
product group, supplemented with a discrete symmetry to enforce the
equality of the gauge couplings.  The simplest theory of this type
that has the same condition on the gauge couplings at the unification
scale as {\sffamily SU(5)} is $\text{G}_\text{TR}\equiv\gtr$
\cite{gtr}.  As we shall see, the minimal trinified model can have as
many as five light Higgs doublets.

\subsection{Minimal Trinification}

We begin by briefly reviewing the minimal trinified model
\cite{Sayre:2006ma,gtr}.  The gauge bosons are assigned to the adjoint
representation, the fermions to $\psi_\Le \oplus \psi_{\qu^c} \oplus
\psi_\qu \equiv \left(1,3,3^\ast\right) \oplus \left(3^\ast,1,3\right)
\oplus \left(3,3^\ast,1\right)$, where
\begin{align*} 
  \psi_\Le & =
  \begin{pmatrix}
    \left(\mathscr{E}\right) & \left(E^c\right) & \left(
      \mathscr{L} \right) \cr \mathscr{N}_1 & e^c & \mathscr{N}_2
  \end{pmatrix}
  , & \psi_{\qu^c} & =
  \begin{pmatrix}
    \mathscr{D}^c \cr u^c \cr \mathscr{B}^c
  \end{pmatrix}
  , &
  \psi_\qu & = \left( \left( -d\quad u \right)\ B
    \vphantom{\frac{B}{B}} \right) .
\end{align*}
The field $(-d,u)$ is the (conjugate of the) usual quark doublet $Q$,
while $B$ is an additional color-triplet, weak-singlet quark.  The
field $u^c$ is the usual up-conjugate quark field, while
$\mathscr{D}^c$ and $\mathscr{B}^c$ have the quantum numbers of the
down-conjugate quark field.  The field $e^c$ is the usual positron
field, and the lepton doublet is a linear combination of $\mathscr{L}$
and $\mathscr{E}$.  The field $E^c$ denotes a lepton doublet with the
opposite hypercharge, and $\mathscr{N}_1$ and $\mathscr{N}_2$ are
sterile (with respect to the standard model) fermions.

$\text{G}_\text{TR}$ is broken by a pair of $\left(1,3,3^\ast\right)$
Higgs fields, which we denote by $\Phi^{1,2}_\Le$,
\begin{align*}
  \Phi_\Le^a & =
  \begin{pmatrix}
    \left(\phi_1^a\right) & \left(\phi_2^a\right) &
    \left(\phi_3^a\right) \cr S_1^a & S_2^a & S_3^a
  \end{pmatrix}
  , & \VEV{\Phi^1_\Le} & =
  \begin{pmatrix}
    u_1 & 0 & 0 \cr 0 & u_2 & 0 \cr 0 & 0 & v_1
  \end{pmatrix}
  , & \VEV{\Phi^2_\Le} & =
  \begin{pmatrix}
    n_1 & 0 & n_3 \cr 0 & n_2 & 0 \cr v_2 & 0 & v_3
  \end{pmatrix}
  .
\end{align*}
For simplicity, we set $v_3=n_i=0$; this does not have any effect on
the qualitative aspects of the model \cite{Sayre:2006ma}.  Both $v_1$
and $v_2$ break $\mathsf{SU(3)_L \times SU(3)_R}$ to
\mbox{$\mathsf{SU(2)_L \times SU(2)_R \times U(1)}$}, but the
$\mathsf{SU(2)_R \times U(1)}$ are different.  Together they break
$\text{G}_\text{TR}$ to the SM.

Of the six Higgs doublets ($\phi_i$, three each in $\Phi^{1,2}_\Le$),
one linear combination is eaten by the gauge bosons that acquire
unification-scale masses. If the remaining five doublets have
weak-scale masses, then gauge-coupling unification results at
$M_\text{U}\simeq 10^{14}$ GeV without supersymmetry.  In general it
would take several fine-tunings to arrange this, so it is an even more
acute form of the usual hierarchy problem.

When $S_3^1$ and $S_1^2$ acquire the vevs $v_1$ and $v_2$,
respectively, $B$ pairs up with
\mbox{$B^c=c_\alpha\,\mathscr{D}^c+s_\alpha\,\mathscr{B}^c$} (where
$\tan\alpha=\tfrac{g_1v_1}{g_2v_2}$ and $s\equiv\sin,\ c\equiv\cos$)
to form a Dirac fermion with a mass at the unification scale,
$m_B^2=g_1^2v_1^2+g_2^2v_2^2$.  Similarly
\mbox{$E=-s_\beta\,\mathscr{E}+c_\beta\,\mathscr{L}$} and $E^c$, where
$\tan\beta=\tfrac{h_1v_1}{h_2v_2}$, form a heavy Dirac state.  The
orthogonal linear combinations of $\mathscr{D}^c$ and $\mathscr{B}^c$
as well as $\mathscr{E}$ and $\mathscr{L}$, which are $d^c$ and $L$,
remain massless.

The electroweak symmetry is broken to $\mathsf{U(1)}_\text{EM}$ when
the two Higgs doublets $\phi^1_1$ and $\phi^1_2$ acquire weak-scale
vevs, $u_1$ and $u_2$.  Then the light fermions acquire masses
\begin{align*}
  m_u & = g_1\, u_2 \ , &
  m_d & = g_1\, u_1\, s_\alpha \ , & 
  m_e & = h_1 u_1\, s_\beta \ , &
  m_{\nu, N_1} & = h_1 u_2 \ , &
  m_{N_2} & \simeq \tfrac{h_1^2 u_1 u_2\, s_\beta}{m_E} \ ,
\end{align*}
where \mbox{$N_1=s_\beta\,\mathscr{N}_1-c_\beta\,\mathscr{N}_2$} and
\mbox{$N_2=-c_\beta\,\mathscr{N}_1-s_\beta\,\mathscr{N}_2$}.  The
results for the fermion masses show that, even in the minimal model,
there is no relation between the masses of the quarks and leptons,
since they depend on five independent parameters
($g_1,h_1,\frac{u_1}{u_2},s_\alpha,s_\beta$).  Thus the minimal
trinification model is sufficient to describe the masses of the quarks
and charged leptons.  At tree level, however, the model yields an
active Dirac neutrino at the weak scale and a sterile Majorana
neutrino at the eV scale; this may be corrected at one loop, via the
``radiative see-saw'' mechanism.

\paragraph{Radiative See-saw Mechanism}
\begin{wrapfigure}{r}{.5\linewidth}
  \begin{flushright}
  \scalebox{0.9}{
    \begin{picture}(200,40)(0,25)
      \Line(0,25)(200,25)
      \Text(0,33)[l]{$N_{1,2},\ \nu$} 
      \Text(80,28)[b]{$q$}
      \Text(120,28)[b]{$q^c$}
      \Text(200,33)[r]{$\nu,\ N_{1,2}$}
      \Line(98,27)(102,23) \Line(98,23)(102,27)
      \DashCArc(100,25)(40,0,180){2}
      \Vertex(60,25)2      \Vertex(140,25)2
      \Line(98,67)(102,63) \Line(98,63)(102,67)
      \Text(100,70)[b]{$\Phi_{\qu^c}\, \Phi_\qu\, \Phi_\text{\sc l}$}
      \Text(60,50)[b]{$\Phi_{\qu^c}$}
      \Text(140,50)[b]{$\Phi_\qu$} 
      \Text(60,20)[t]{$\psi_\qu \psi_\text{\sc l} \Phi_{\qu^c}$}   
      \Text(140,20)[t]{$\psi_\text{\sc l} \psi_{\qu^c} \Phi_\qu$} 
    \end{picture}
  }
  \end{flushright}
\end{wrapfigure}
Large radiative contributions to the masses of the sterile neutrinos
occur due to the coupling of the neutral fermions to color-triplet
Higgs bosons, $\Phi_\qu$ and $\Phi_{\qu^c}$, and the cubic couplings
of the Higgs fields.  These interactions may be used to construct the
one-loop diagram, displayed in the adjoining figure.  This diagram is
dominated by the quark that acquires a unification-scale mass, namely
the heavy $B$ quark.  It yields large one-loop contributions such that
the effective neutrino mass matrix, in the $\left( \nu, N_1, N_2
\right)$ basis, is
\begin{align*}
  M_N^\text{1-loop} \simeq
  \begin{pmatrix}
    0 & -h_1 u_2 & 0 \cr -h_1 u_2 & s_{\alpha-\beta} c_\beta\, g^2 F_B
    & \left( s_{2\beta} s_\alpha - c_\alpha \right) g^2 F_B \cr 0 &
    \left( s_{2\beta} s_\alpha - c_\alpha \right) g^2 F_B &
    c_{\alpha-\beta} s_\beta\, g^2 F_B
  \end{pmatrix} ,
\end{align*}
where $F_B$ denotes the loop integral.  This matrix has two
eigenvalues,
\begin{align*}
  m_{N_{1,2}} & \sim g^2 F_B \ , & m_{\nu} \sim \tfrac{h_1^2
    u_2^2}{g^2 F_B} \ .
\end{align*}
Thus the two sterile neutrinos acquire unification-scale masses at one
loop, while the active neutrino acquires a ``radiative see-saw''
Majorana mass.  In order to obtain the correct values for the tau and
top masses, we expect $h_1\sim 0.1$, $g\sim 1$ and $u_2\sim 100$ GeV.
Since $F_B\simeq M_\text{U}/(4\pi)^2$, the mass of the light neutrino
is then ${\cal O}\left(0.1\,\text{eV}\right)$, consistent with the
experimental constraints \cite{nu-masses}.

There is also a one-loop diagram that couples $\nu$ to $N_{1,2}$ but
with the heavy $B$ quark replaced by $d$.  This diagram is comparable
to the tree-level Dirac mass $h_1u_2$, so it does not qualitatively
change the radiative see-saw mechanism.

The radiative see-saw mechanism is absent in models with weak-scale
supersymmetry, since the one-loop contributions are reduced to
$\mathcal{O}\left(1\,\text{TeV}\right)$, but it is present if the mass
difference between scalars and fermions is comparable to the
grand-unified scale, as may be the case in split supersymmetry
\cite{split-susy}.

\paragraph{Neutrino Hierarchy}
From the discussion above, we note that the neutrino hierarchy is
related to the couplings in the quark sector.  Since the one-loop
contributions to the neutrino masses are proportional to the fermion
mass in the loop, those with the heaviest quark, $B_3$, are dominant.

For our discussion, we assume that the matrices $g$ have the form
\cite{Lola:1999un}
\begin{align} \label{eq:up-lola}
  g & \sim
  \begin{pmatrix}
    \epsilon^4 & \epsilon^3 & \epsilon^3 \cr \epsilon^3 & \epsilon^2 &
    \epsilon^2 \cr \epsilon & 1 & 1
  \end{pmatrix}
  , \quad \epsilon^2 \sim \frac{m_c}{m_t} \ , & \mathcal{G} & = g_{3i}
  g_{j3} + g_{i3} g_{3j} \sim
  \begin{pmatrix}
    \epsilon^4 & \epsilon^3 & \epsilon \cr \epsilon^3 & \epsilon^2 & 1
    \cr \epsilon & 1 & 1
  \end{pmatrix} 
  .
\end{align}
To obtain a hierarchical structure for the up and down quarks, both
$g_1$ and $g_2$ will generally be hierarchical, so we expect the $B$
quarks to have a similar hierarchy.  Then the three-generational mass
matrix for the sterile neutrinos (both $N_1$ and $N_2$) has the
eigenvalues $m^N_3 \sim m^N_2 \sim F_{B_3} \sim 10^{12}$ GeV and
$m^N_1 \sim \epsilon^4 F_{B_3} \sim 10^8$ GeV; the latter is
comparable to the mass of the lightest sterile neutrino in thermal
leptogenesis.

The eigenvalues of the light neutrinos are proportional to $h^2/g^2$
due to the common loop-integral, where $g^2$ is given by the third
column of the symmetric matrix $\mathcal{G}$ in
Eq.~(\ref{eq:up-lola}).  Since the hierarchy of $g^2$ is weak, the
neutrino hierarchy is determined by the hierarchy of $h$ and we find
either quasi-degenerate masses or a normal hierarchy.

\paragraph{Proton Decay}
The gauge interactions conserve baryon number, and therefore do not
mediate proton decay.  Instead, proton decay is mediated by the
colored Higgs fields.  These dimension-six operators are suppressed by
the small Yukawa couplings.  Hence, the flavor non-diagonal decay is
dominant.

We can estimate the lifetime to be $\tau \simeq \left(\frac{1}{g\,
    h}\right)^2 \times 10^{28}$ years; for details, see
Ref.~\cite{Sayre:2006ma}.  Using $g$ as given in
Eq.~(\ref{eq:up-lola}), we find the channel $p\to\bar\nu K^+$ be the
dominant decay mode, followed by $p\to\mu^+ K^0$.  This result is
similar to those in models with weak-scale supersymmetry, where the
decay is dominated by dimension-five operators
\cite{Wiesenfeldt:2004qa}.  The decay might be observable in future
experiments which aim to reach a lifetime of $10^{35-36}$ years
\cite{future-exp}.  In the split supersymmetry scenario, however,
proton decay is unobservable due to the higher unification scale and
the large sfermion masses.

\paragraph{Conclusion}
The minimal trinified model is an interesting and viable candidate for
non-supersymmetric unification.  The breaking is achieved by only two
$\left(1,3,3^\ast\right)$ Higgs fields, which include five Higgs
doublets.  Unlike other grand-unified theories, this model is able to
correctly describe the fermion masses and mixing angles without the
need to introduce intermediate scales, additional Higgs fields, or
higher-dimensional operators.

Light, active neutrinos are naturally generated at one loop via the
radiative seesaw-mechanism.  The additional matter, which is either
vectorlike or sterile, is superheavy with masses above $10^8$\,GeV.
Thus no additional particles are present at the weak scale.

Proton decay is mediated by colored Higgs bosons with the dominant
decay modes potentially observable in future experiments.  Therefore
the different types of models would be distinguished by the presence
or absence of supersymmetric particles and the number of Higgs
doublets at the weak scale, together with the observation of specific
proton decay modes \cite{Wiesenfeldt:2004qa}.  The smoking gun for
minimal trinification would be the discovery of five Higgs doublets at
the weak scale and the observation of proton decay into final states
containing kaons.

\paragraph{Acknowledgments}
It is a pleasure to thank J.~Sayre and S.~Willenbrock for
collaboration and the organizers of SUSY06 for an enjoyable
conference.

\bibliographystyle{aipproc}

\begin{thebibliography}{9}
  
\bibitem{unification-higgs} 
  S.~L.~Adler, Phys.\ Rev.\ D {\bf 59}, 015012 (1999) [Erratum-ibid.\ 
  D {\bf 59}, 099902 (1999)];
  ibid. D {\bf 60}, 015002 (1999);
  S.~Willenbrock, Phys.\ Lett.\ B {\bf 561}, 130 (2003).
  
\bibitem{Sayre:2006ma} J.~Sayre, S.~Wiesenfeldt and S.~Willenbrock,
  Phys.\ Rev.\ D {\bf 73}, 035013 (2006).

\bibitem{gtr} 
  Y.~Achiman and B.~Stech, in {\sl Advanced Summer Institute on New
    Phenomena in Lepton and Hadron Physics}, eds.  D.~E.~C.~Fries and
  J.~Wess (Plenum, New York, 1979);
  S.~L.~Glashow, in {\sl Fifth Workshop on Grand Unification},
  eds.~K.~Kang, H.~Fried, and P.~Frampton (World Scientific,
  Singapore, 1984), p.~88;
  K.~S.~Babu, X.~G.~He and S.~Pakvasa, Phys.\ Rev.\ D {\bf 33}, 763
  (1986).

\bibitem{nu-masses} 
  S.~M.~Bilenky, C.~Giunti, J.~A.~Grifols and E.~Masso, Phys.\ Rept.\ 
  {\bf 379}, 69 (2003);
  A.~Strumia and F.~Vissani, Nucl.\ Phys.\ B {\bf 726}, 294 (2005).
  
\bibitem{split-susy} 
  N.~Arkani-Hamed and S.~Dimopoulos,
  JHEP {\bf 0506}, 073 (2005);
  G.~F.~Giudice and A.~Romanino, Nucl.\ Phys.\ B {\bf 699}, 65 (2004)
  [Erratum-ibid.\ B {\bf 706}, 65 (2005)].

\bibitem{Lola:1999un} S.~Lola and G.~G.~Ross, Nucl.\ Phys.\ B {\bf
    553}, 81 (1999).
  
\bibitem{Wiesenfeldt:2004qa} S.~Wiesenfeldt, Mod.\ Phys.\ Lett.\ A
  {\bf 19}, 2155 (2004).
      
\bibitem{future-exp} 
  J.~G.~Learned, {\slshape 9th International Symposium on Neutrino
    Telescopes, Venice, Italy, 6-9 Mar 2001}, {\tt
    http://axpd24.pd.infn.it/conference2001/proceedings/Learned.ps};
  \\
  A.~Rubbia, hep-ph/0407297.

\end{thebibliography}

\end{document}